# Comparison of Magnetic Field Characteristics among 3 Tesla MRI Scanners: An Experimental Measurement Study

FRANCESCO GIRARDELLO[1], MARIA ANTONIETTA D'AVANZO[2], MASSIMO MATTOZZI[2], VICTORIAN MICHELE FERRO[3], GIUSEPPE ACRI[3, a] AND VALENTINA HARTWIG[1, a]

[1] CNR, Institute of Clinical Physiology
56124 Pisa, ITALY.
[2] INAIL, Department of Occupational and Environmental Medicine
Via Fontana Candida 1, 00078 Rome, ITALY.
[3] Dipartimento di Scienze Biomediche, Odontoiatriche e delle Immagini Morfologiche e Funzionali,
Università degli Studi di Messina, 98125 Messina, ITALY.
a) Corresponding authors: giuseppe.acri@unime.it
valentina.hartwig@cnr.it

*Abstract:* - Magnetic resonance imaging (MRI) scanners have advanced significantly, with a growing use of high-field 3 T systems. This evolution gives rise to safety concerns for healthcare personnel working in proximity to MRI equipment. While manufacturers provide theoretical Gauss line projections, these are typically derived under ideal open-environment conditions and may not reflect real-world installations. For this reason, identical MRI models can produce markedly different fringe field distributions depending on shielding and room configurations. The present study proposes an experimental methodology for the mapping of the fringe magnetic field in the vicinity of three 3 T MRI scanners. Field measurements were interpolated to generate three-dimensional magnetic field maps. A comparative analysis was conducted, which revealed notable differences among the scanners. These differences serve to highlight the influence of site-specific factors on magnetic field propagation.

## 1 Introduction

Magnetic resonance imaging (MRI) is one of the most widely utilized diagnostic imaging modalities [1]. In addition, its applications are manifold, encompassing the domains of veterinary radiology, paleontology, forensic imaging, and industrial engineering. The issue of MRI safety is of particular significance and cannot be disregarded; it requires constant attention and adherence to safety protocols by MR technologists and staff. Firstly, the intense static field (from 0.5 to 11.7 T), which is constantly present in the superconductor magnet, gives rise to significant safety concerns. This is particularly the case with regard to the projectile risks associated with ferromagnetic objects drawn to the scanner. Furthermore, the static magnetic field has the capacity to interfere with medical devices that incorporate magnetic components, thereby causing malfunctions. In MRI workplaces, movement through the spatial gradient of the static field (fringe field) acts as a time-varying magnetic field (motion-induced TvMF) [2], [3], [4], inducing a voltage in electrically conductive materials, such as biological tissues, according to Faraday's law [5].

It has been demonstrated that rapid movements of the body induce a significant electric field in the tissue, which has been shown to induce a number of physiological symptoms, including, but not limited to, headache, nausea, vertigo, phosphenes, numbness and tingling, loss of proprioception, and balance. In order to perform a rigorous risk assessment within an MRI scanner, it is imperative to possess comprehensive and precise knowledge regarding the magnetic field that is distributed throughout the scanner [6], [7]. Scanner manufacturers generally provide isogauss line plans, which frequently lack the requisite level of detail to adequately represent the fringe field. Specifically, the zone closest to the gantry, which exhibits the highest spatial gradient value, is not detailed enough [8], [9]. In addition, the isogauss line plans provided by manufacturers are generally associated with static fields in the absence of supplementary shielding [10]. Depending on the environmental factors present in the installation location, scanners of the same type and built by the same manufacturer may exhibit a variation in shielding. It is imperative to ascertain the fringe field post-installation in the operating environment of the specific MRI scanner.

The primary aim of this study is to investigate and quantify the variations in magnetic field distributions generated by 3 Tesla MRI scanners installed in different clinical facilities. The objective is to provide valuable insights into MRI safety, with a particular focus on magnetic field leakage, thereby facilitating a more profound comprehension of MRI environments. The study's outcomes will be of significant value to MRI facilities seeking to optimize room design and safety measures, particularly in light of the ongoing growth in demand for higher-powered MRI systems.

## 2 Methodology

The experimental analysis conducted encompasses three distinct Italian healthcare facilities: the FTGM Ospedale del Cuore in Massa, Ospedale San Marco in Catania, and Ospedali Riuniti in Reggio Calabria. While all three installations feature 3 Tesla MRI systems, it is noteworthy that the scanner at FTGM Ospedale del Cuore is manufactured by a different vendor compared to the other two facilities, providing an opportunity to evaluate both inter-manufacturer and intra-manufacturer variability.

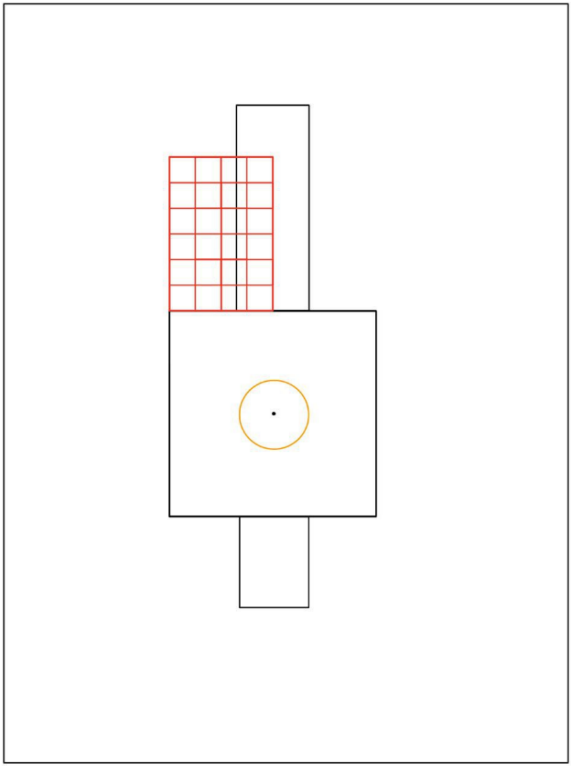

Fig. 1: Schematic representation of an MRI room axial view. In red, the $10 \times 10$ cm grid where measurements of the magnetic field have been taken.

To comprehensively characterize the magnetic fields across the three clinical MRI scanners, systematic measurements of the magnetic field magnitude and its spatial components were conducted throughout each MRI suite using a commercial HP-01 magnetometer (Narda Safety Test Solutions, Savona, Italy). The measurement protocol employed a well-consolidated multi-plane approach at three anatomically relevant heights: a lower plane corresponding to lower abdomen/waist level, an intermediate plane at thoracic/heart level, and an upper plane at head/eyes level. At each measurement plane, we collected a minimum of 120 data points distributed across a $10 \times 10$ cm grid matrix positioned adjacent to the gantry on the right side of the patient table (red box in Figure 1). This high-density spatial sampling approach enabled detailed mapping of the magnetic field distribution with sufficient resolution to identify relevant spatial gradients. The complete data acquisition methodology adhered to the validated protocol described in detail by Hartwig et al. (9), ensuring procedural consistency and reproducibility across all three clinical facilities.

The static magnetic field of three different MRI environments was assessed: Table 1 reports the main characteristics of the chosen sites, together with the relative locations.

Table 1: Main characteristics of MRI sites

| ID | Scanner | Location | B0 (T) | Shielding |
|---|---|---|---|---|
| **MRI-1** | Total body clinical MRI scanner – Horizontal oriented | FTGM Ospedale del Cuore (Massa, Italy) | 3 | Active |
| **MRI-2**[*] | Total body clinical MRI scanner – Horizontal oriented | Ospedale San Marco (Catania, Italy) | 3 | Active |
| **MRI-3**[*] | Total body clinical MRI scanner – Horizontal oriented | Ospedali Riuniti (Reggio Calabria, Italy) | 3 | Active |

*same manufacturer

## 3 Data Processing

All computational analyses described in this study were performed using custom-developed scripts in MATLAB® R2020b (MathWorks, Inc., Natick, MA, USA).

To create a comprehensive magnetic field map of the entire MRI suite, we extended discrete measurement points across the complete cross-sectional area of each room. The spatial distribution of the magnetic field was modelled using parametric fitting applied to each row and column of the experimental data matrix. This approach enabled spatial extrapolation of the measured data along both the x and z axes, with approximately 20 fits performed per measurement plane, allowing for field prediction throughout the entire room.

The fitting procedure was implemented using a nonlinear least-squares optimization algorithm that determines optimal parameters by minimizing the sum of squared residuals between model and experimental data. To ensure convergence and robustness, a custom routine was developed for computing initial parameter values, designed to generate plausible starting points based on data amplitude and spatial scale. The fitting process required systematic evaluation of a set of parametric functions specifically selected to capture the complexity of spatial variations in the magnetic field. Approximately ten types of functions were implemented, consisting of a combination of polynomial and exponential forms with 5 and 6 free parameters. Representative functions include:

$$f_{\exp}(\boldsymbol{b}, x) = b_1 e^{-b_2 x} + b_3 e^{-b_4 x} + b_5 e^{-b_6 x} \quad (1)$$

$$f_{\text{poly}}(\boldsymbol{b}, x) = b_1 e^{-b_2 x} \cdot (b_3 x^2 + b_4 x + b_5) \quad (2)$$

$$f_{\text{exp-poly}}(\boldsymbol{a}, x) = a_1 e^{-|a_2|x - |a_3|x^2} \cdot (1 + a_4 x + a_5 x^2) \quad (3)$$

The selection of this extensive repertoire of complex functions was motivated by two fundamental considerations. First, the need to extend experimental data both along the x-axis and z-axis for different heights results in significant variations in magnetic field profiles, requiring flexibility in mathematical modelling. Second, the approach aims to develop a generalizable characterization methodology, potentially applicable to any hospital configuration with different geometries and MRI scanner types.

The optimization procedure consists of two distinct phases. In the first phase, for each candidate function, the algorithm determines optimal parameters by minimizing the sum of squared residuals. In the second phase, to compare the performance of different already-optimized functions, the reduced chi-squared is calculated for each:

$$\chi^2_{red} = \frac{\Sigma \left(\frac{residual_i}{B_i}\right)^2}{n - p} \quad (4)$$

where *residual* represents the difference between the measured value and the model prediction for the i-th point, $B_i$ is the measured magnetic field value at point *i*, *n* is the total number of experimental points, and *p* is the number of parameters in the fitting function. The function producing the $\chi^2_{red}$ value closest to unity is selected as the optimal model for that specific spatial region. This two-level approach ensures both parameter optimization for each individual function and objective selection of the most appropriate mathematical model for accurate local magnetic field representation.

To complete the spatial mapping, an additional interpolation phase was implemented for the outermost regions. The intrinsic symmetries of the magnetic field were then leveraged to achieve a complete mapping of the entire three-dimensional space. This approach allowed us to transform discrete measurement points into comprehensive volumetric field maps while maintaining physical consistency throughout the mapped volume, providing unprecedented insight into the spatial distribution of fringe fields in clinical MRI installations.

Generation of a comprehensive three-dimensional fringe field map within the MRI suite necessitates the acquisition and interpolation of a minimum of three complete measurement planes. The precision of the volumetric interpolation process exhibits direct proportionality to the number of acquired planes. To optimize the volumetric reconstruction, we exploited the inherent symmetry of the magnetic field with respect to the y-axis, effectively doubling the number of planes available for interpolation.

Isogauss plot lines, as provided by the manufacturer, demonstrate the spatial distribution of the static field relative to the magnet isocentre from the lateral, superior, and anterior perspectives. In order to sustain this standard perspective, the estimated fringe field values are depicted as contour maps on the three reference planes. Furthermore, the fringe field contour map at the head height is represented in order to highlight the exposure condition in such a position.

The uncertainty analysis for this study considers three distinct categories of data points, each associated with different error sources. For directly measured points, the uncertainty is solely attributed to instrumental error. For points obtained through fitting procedures, the total uncertainty is calculated as the quadratic combination of instrumental error and fitting-related statistical uncertainty. Finally, for interpolated points, the overall uncertainty incorporates the quadratic sum of instrumental error, fitting uncertainty, and interpolation error.

All measurements were acquired using the NARDA HP01 gaussmeter, which exhibits an experimental uncertainty of 1% ($\sigma_{exp}$) under the operating conditions employed in this study. The device was regularly calibrated according to the manufacturer's guidelines. For fitted data points, the total uncertainty ($\sigma_{fit}$) was calculated using standardized residuals analysis [11], providing an uncertainty estimate based on fit quality and residual distribution.

The interpolation error was quantified following theoretical principles of uncertainty propagation in interpolated scales [12]. The interpolation error was calculated as:

$$\sigma_{\text{interp}} = \sigma_i + \alpha \times d_{\min} \quad (5)$$

where $\sigma_i$ represents the intrinsic uncertainty of the scattered interpolant method (3%), α is the distance coefficient (0.05% per cm), and $d_{\min}$ is the minimum distance from measurement points.

The overall uncertainty was determined by a quadratic combination of all error components:

$$\sigma_{\text{totale}} = \sqrt{\sigma_{\exp}^2 + \sigma_{\text{fit}}^2 + \sigma_{interp}^2} \quad (6)$$

where $\sigma_{exp}$ represents the experimental uncertainty associated with the measurement instrument, $\sigma_{fit}$ refers to the uncertainty introduced by the fitting procedure, and $\sigma_{interp}$ accounts for the interpolation procedure performed. This comprehensive error analysis ensures accurate uncertainty quantification throughout the mapped volume, reflecting the fundamental principle that interpolation uncertainty increases with distance from sampling points.

Finally, the rate of change of the magnetic field as a function of position around the MRI scanner, i.e., the spatial field gradient (SFG), was calculated using the partial derivatives of the interpolated total magnetic field. The gradient was determined by applying the differential operator to the total field map with a step size of 1 cm. The magnitude was calculated as the square root of the sum of squares of the components along the axes relative to the reference plane and converted to Tesla/meter (T/m). Furthermore, the SFG was compared among the three MRI scanners to evaluate differences in the spatial distribution of field gradients.

## 4 Results

The results presented refer to the three scanners indicated in Table 1. The presentation of results will be organized as follows: a first section dedicated to images related to the magnetic field, a second section to the magnitude of its spatial gradient, and a final summary section with four tables (two for the magnetic field and two for the gradient).

Before proceeding with the analysis of the obtained results, it is necessary to evaluate the quality of the fitting models applied to the experimental data. For each hospital, approximately 60 fits were performed to extend the data along the x and z axes at three different heights. Table 2 summarizes the reduced χ² values calculated for the different experimental configurations, specifying for each hospital and for each measurement plane: the number of fits performed, the mean value of the reduced χ², as well as the maximum and minimum values recorded.

Figures 2 to 4 represent the data related to pairwise comparisons of the three scanners for perpendicular planes passing through the isocenter,

with multiple isogauss lines fixed to specific values
(0.1 T, 0.5 T, 1 T, 1.5 T, 2 T, 2.5 T, 3 T).

Table 2: Fitting model parameters for the fringe field map computation

|        | Height (cm) | N° Fit | $\chi^2$ mean | $\chi^2$ min | $\chi^2$ max |
|--------|-------------|--------|---------------|--------------|--------------|
| MRI-1  | 95          | 19     | 0.94          | 0.55         | 1.50         |
| MRI-1  | 138         | 20     | 0.96          | 0.44         | 1.28         |
| MRI-1  | 160         | 20     | 0.90          | 0.20         | 1.42         |
| MRI-2  | 100         | 22     | 0.98          | 0.38         | 1.62         |
| MRI-2  | 130         | 19     | 0.95          | 0.27         | 1.43         |
| MRI-2  | 160         | 18     | 0.98          | 0.29         | 2.09         |
| MRI-3  | 100         | 22     | 0.94          | 0.05         | 1.44         |
| MRI-3  | 130         | 19     | 0.84          | 0.20         | 1.42         |
| MRI-3  | 160         | 19     | 1.03          | 0.49         | 1.91         |

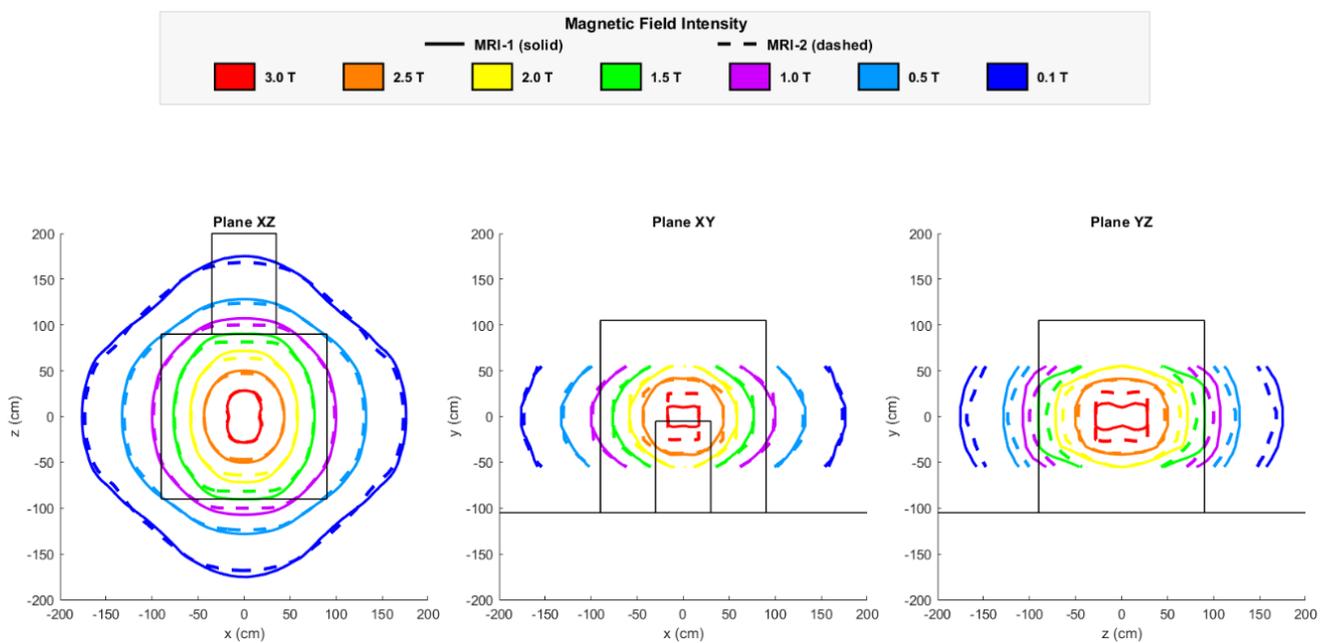

Fig. 2: Isogauss contours of the three perpendicular planes passing through the scanner isocenter with fixed contour lines. Comparison between MRI-1 machine (solid lines) and MRI-2 machine (dashed lines).

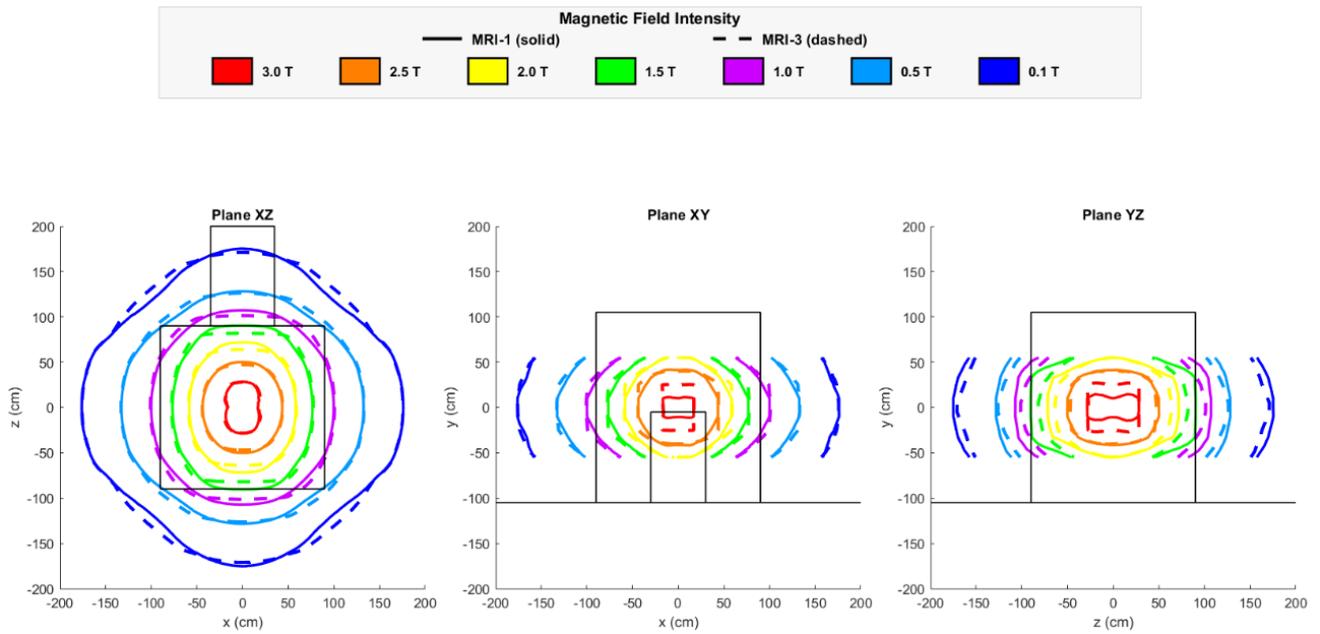

Fig. 3: Isogauss contours of the three perpendicular planes passing through the scanner isocenter with fixed contour lines. Comparison between MRI-1 machine (solid lines) and MRI-3 machine (dashed lines).

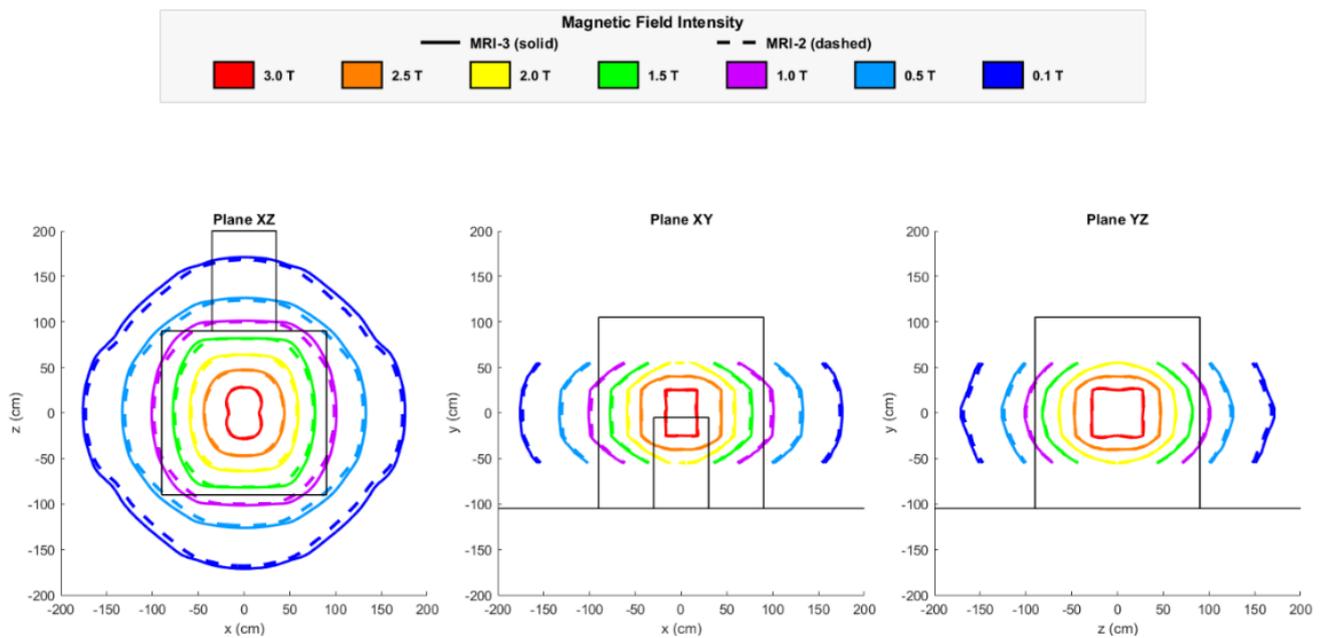

Fig 4: Isogauss contours of the three perpendicular planes passing through the scanner isocenter with fixed contour lines. Comparison between MRI-2 machine (solid lines) and MRI-3 machine (dashed lines).

Figure 5 shows the isogauss lines for the planes parallel to the floor at y = 55 and y = 0 cm. These two graphs are relevant because for a person 170 cm tall, they represent approximately the height of the head/eyes (y = 55 cm, that is at 160 cm from the floor) and waist (y = 0 cm, that is at 105 cm from the floor) [13].

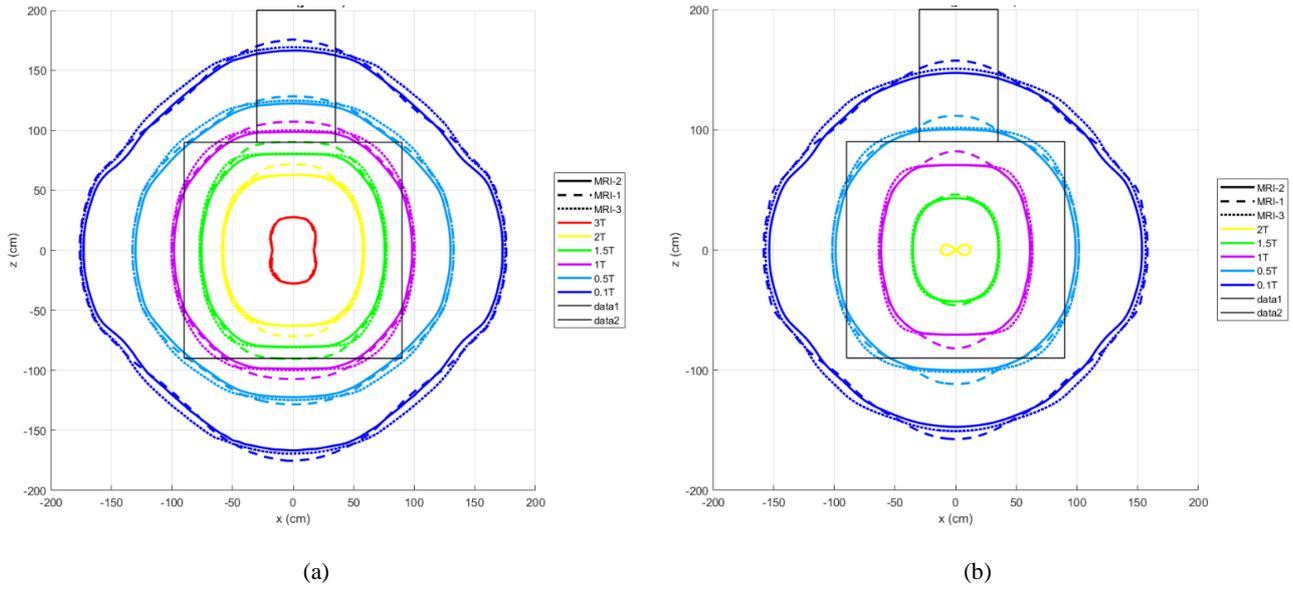

(a)

(b)

Fig. 5: Isogauss contours for the three scanners in the XZ plane at y = 0 cm - fixed contour lines (a). Isogauss contours for the three scanners in the XZ plane at y = 55 cm - fixed contour lines (b).

In addition to visualizing the isogauss lines for the three facilities, it is interesting to also analyse the magnetic field differences. Indeed, Figures 6, 7, 8, 9, 10 show the magnetic field difference for the various hospitals compared pairwise. As before, the first three figures relate to planes passing through the isocentre, while the last two relate to planes parallel to the floor. In all figures, a red dot represents the location where the difference between the two compared scanners is greatest; the numerical value of this difference is reported in the subsequent tables.

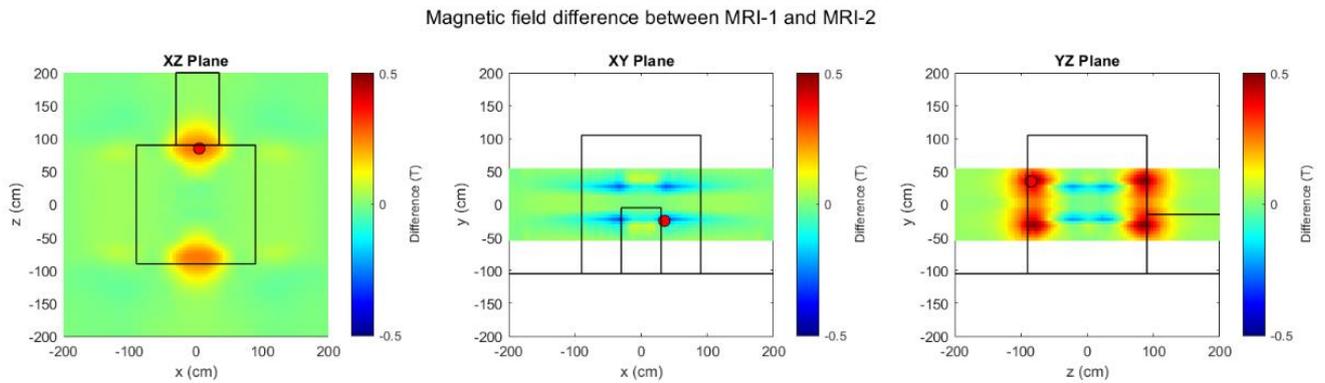

Fig. 6: Magnetic field difference between MRI-1 and MRI-2 scanners for the three perpendicular planes passing through the isocenter. The red dot represents the location where the values show the greatest deviation.

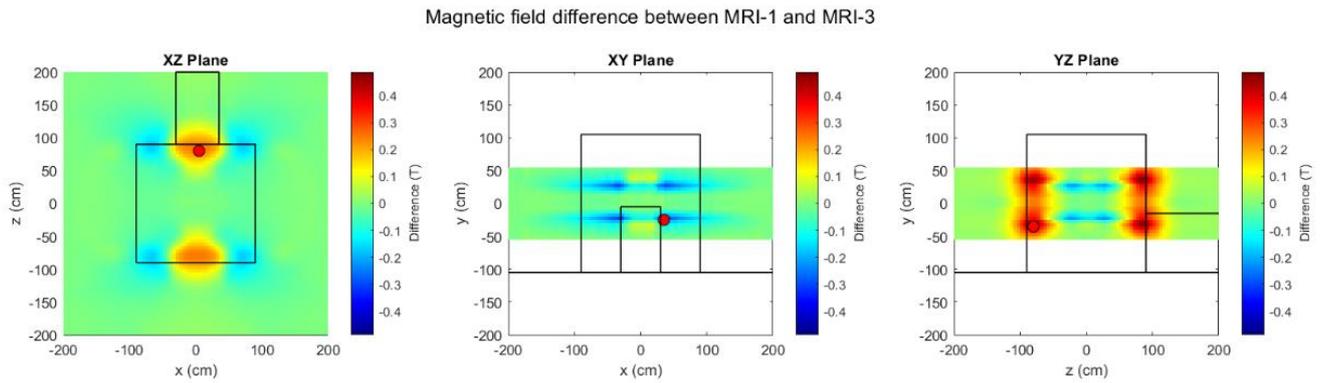

Fig. 7: Magnetic field difference between MRI-1 and MRI-3 scanners for the three perpendicular planes passing through the isocenter. The red dot represents the location where the values show the greatest deviation.

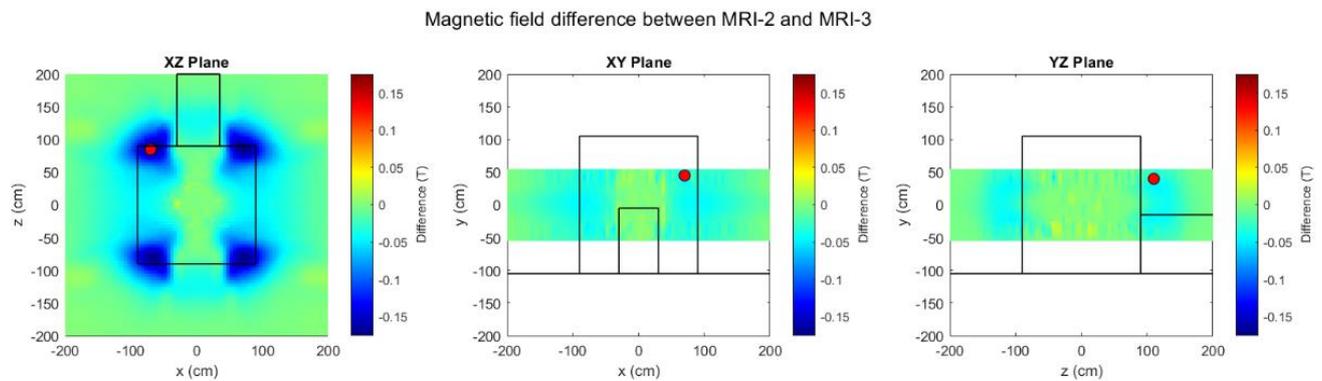

Fig. 8: Magnetic field difference between MRI-2 and MRI-3 scanners for the three perpendicular planes passing through the isocenter. The red dot represents the location where the values show the greatest deviation.

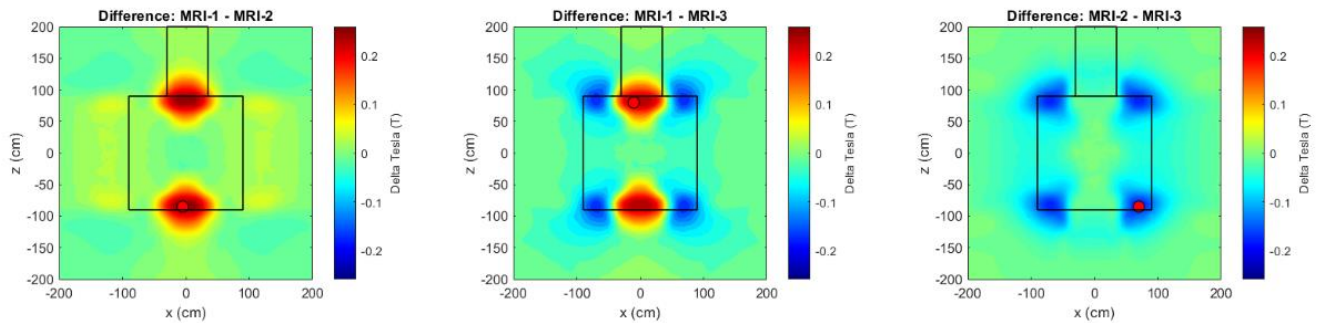

Fig. 9: Magnetic field difference at height y = 0 cm for all possible combinations of the three scanners. The red dot represents the location where the values show the greatest deviation.

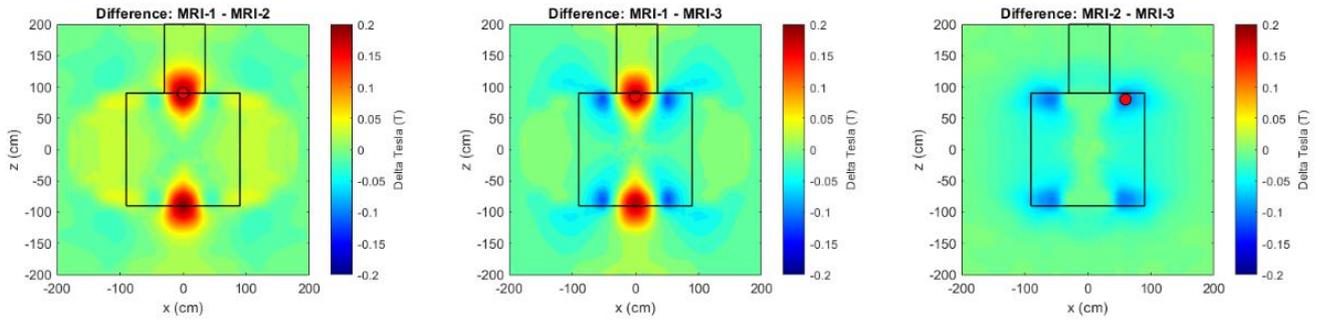

Fig. 10: Magnetic field difference at height y = 55 cm for all possible combinations of the three scanners. The red dot represents the location where the values show the greatest deviation.

The subsequent analysis has been relative to the spatial gradient magnitude for the aforementioned planes that have been previously discussed. Figures 11, 12, and 13 illustrate comparisons between isogradient lines of the three orthogonal planes at the isocentre for the three hospitals. The gradient values in the internal area of the scanner, shown as a dotted line circle, are not reported.

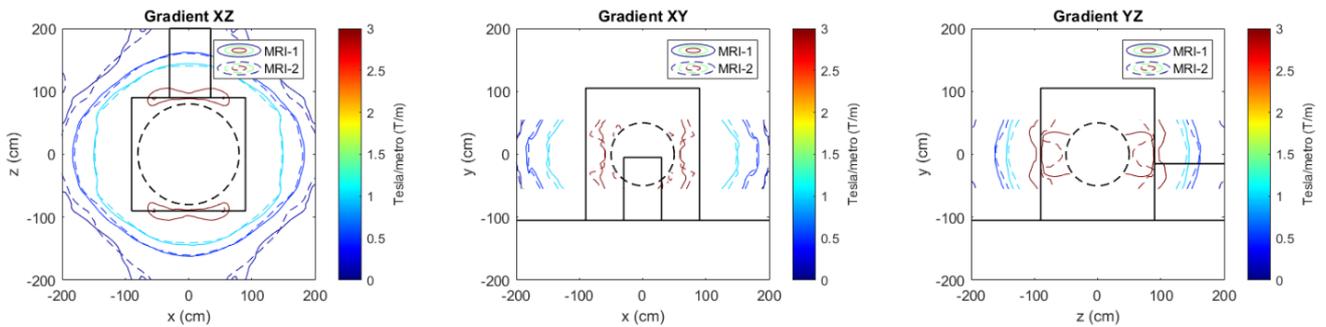

Fig. 11: Comparison of isogradient between MRI-1 and MRI-2 scanners for the three orthogonal planes passing through the isocenter.

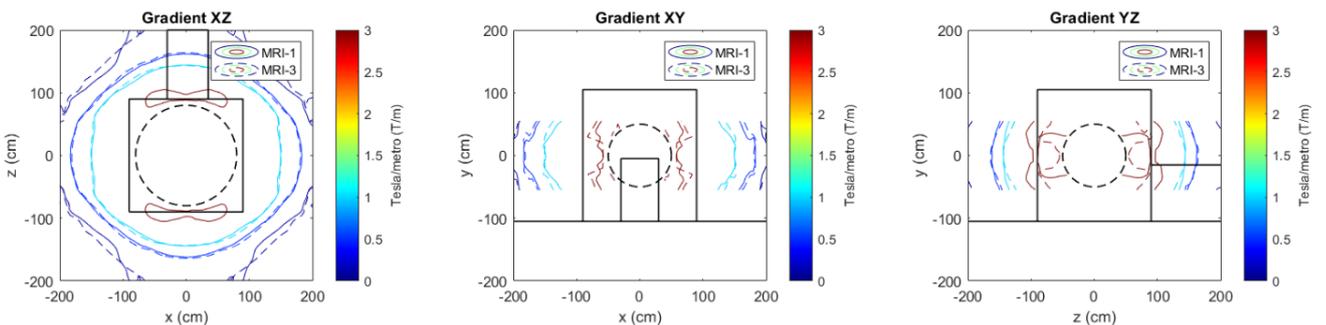

Fig. 12: Comparison of isogradient between MRI-1 and MRI-3 scanners for the three orthogonal planes passing through the isocenter.

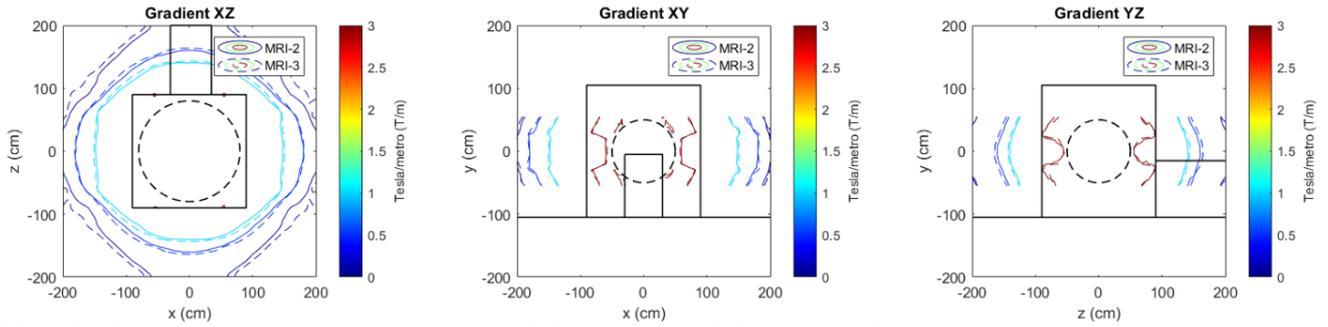

Fig. 13: Comparison of isogradient between MRI-2 and MRI-3 scanners for the three orthogonal planes passing through the isocenter.

Figure 14 shows the isogradient lines for the two planes parallel to the floor (y = 0 cm and y = 55 cm).

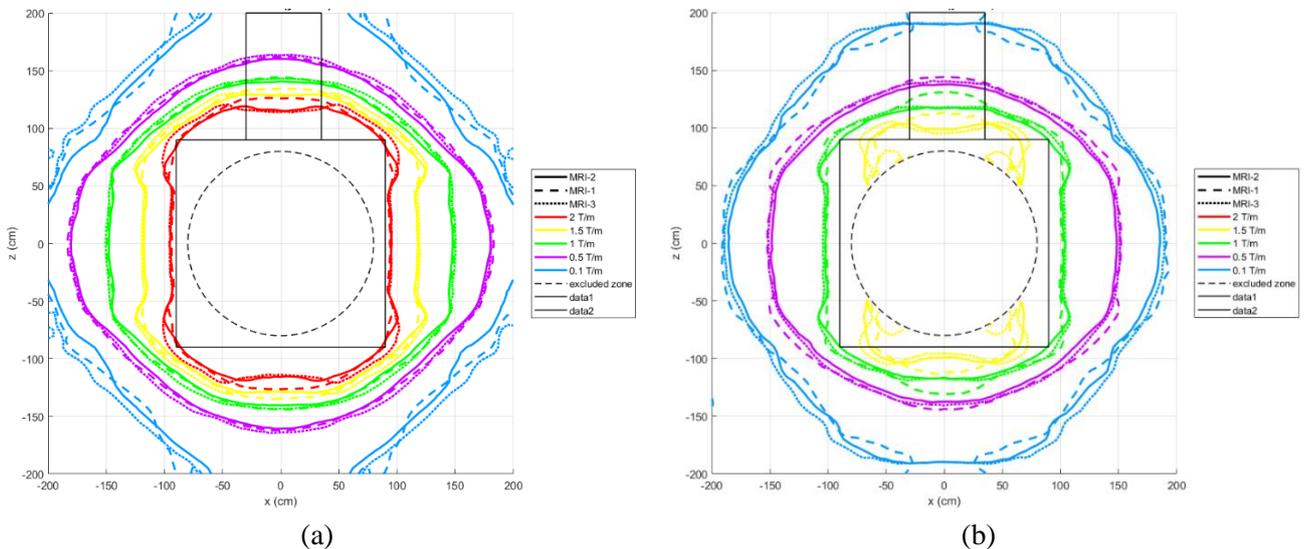

(a)                  (b)

Fig. 14: Isogradient of the three scanners considered with fixed contour lines for a plane parallel to the floor at height y = 0 cm (a). Isogradient of the three scanners considered with fixed contour lines for a plane parallel to the floor at height y = 55 cm (b).

Tables 3 and 4 regarding the magnetic field and spatial gradient values relative to three orthogonal planes passing through the isocentre, and the plane XZ at a height of 160 cm from the floor (y = 55 cm). For each plane, the point at which the maximum difference between two scanners is observed is reported, along with the value of the first scanner, the value of the second scanner, and the difference between them. This analysis was conducted for all three scanner combinations, both for the magnetic field (see Table 3) and for the gradient (see Table 4). The tables provide a quantitative summary of the points of maximum divergence between scanners, thus allowing direct comparison of measured values and relative differences for each pair of devices in the planes of clinical interest.

Table 3: Summary of maximum magnetic field difference points between the various scanners (value ± uncertainty)

| Plane | Coordinates (cm) | Field Value (T) | Field Value (T) | Difference (T) |
|---|---|---|---|---|
| *MRI-1 vs MRI-2* | | *MRI-1* | *MRI-2* | |

| Plane | Coordinates (cm) | | | |
|---|---|---|---|---|
| XZ | x = 5; y = 0; z = 85 | 1.66 ± 0.04 | 1.40 ± 0.06 | 0.26 ± 0.07 |
| XY | x = 35; y = -25; z = 0 | 2.51 ± 0.42 | 2.81 ± 0.47 | -0.30 ± 0.63 |
| YZ | x = 0; y = 35; z = -85 | 1.58 ±0.04 | 1.08 ± 0.05 | 0.50 ± 0.06 |
| *XZ | x = 0; y = 55; z = -85 | 0.95 ± 0.03 | 0.75 ± 0.04 | 0.20 ± 0.06 |
| | | | | |
| *MRI-1 vs MRI-3* | | *MRI-1* | *MRI-3* | |
| XZ | x = 5; y = 0; z = 80 | 1.80 ± 0.05 | 1.55 ± 0.14 | 0.35 ± 0.146 |
| XY | x = 35; y = -25; z = 0 | 2.51 ± 0.42 | 2.81 ± 0.47 | -0.30 ± 0.63 |
| YZ | x = 0; y = -35z = -80 | 1.71 ± 0.06 | 1.23 ± 0.15 | 0.48 ± 0.15 |
| *XZ | x = 0; y = 55; z = -85 | 0.94 ± 0.03 | 0.76 ± 0.04 | 0.18 ± 0.05 |
| | | | | |
| *MRI-2 vs MRI-3* | | *MRI-2* | *MRI-3* | |
| XZ | x = -70; y = 0; z = 85 | 0.85 ± 0. 02 | 1.02 ± 0.03 | -0.17 ± 0.15 |
| XY | x = 70; y = 55; z = 0 | 1.11 ± 0.14 | 1.15 ± 0.14 | -0.04 ± 0.62 |
| YZ | x = 0; y = 40; z = 110 | 0.50 ± 0.02 | 0.54 ± 0.03 | -0.04 ± 0.01 |
| *XZ | x = -55; y = 55; z = -80 | 0.64 ± 0.02 | 0.74 ± 0.02 | -0.10 ± 0.02 |

*relative to head/eye level

Table 4: Summary of maximum magnetic gradient difference points between the various scanners (value ± uncertainty)

| Plane | Coordinates (cm) | Gradient Magnitude (T/m) | Gradient Magnitude (T/m) | Difference (T/m) |
|---|---|---|---|---|
| *MRI-1 vs MRI-2* | | *MRI-1* | *MRI-2* | |
| XZ | x = 30; y = 0; z = 95 | 3.34 ± 0.09 | 2.68 ± 0.07 | 0.66 ± 0.11 |
| XY | x = 80; y = - 30; z = 0 | 2.44 ± 0.32 | 3.21 ± 0.41 | -0.77 ± 0.52 |
| YZ | x = 0; y = -30; z = -75 | 1.93 ± 0.05 | 3.57 ± 0.23 | -1.64 ± 0.23 |
| *XZ | x = 0; y = 55; z = 120 | 1.38 ± 0.04 | 0.92 ± 0.04 | 0.46 ± 0.06 |
| | | | | |
| *MRI-1 vs MRI-3* | | *MRI-1* | *MRI-3* | |
| XZ | x = -40; y = 0; z = -95 | 3.36 ± 0.10 | 2.61 ± 0.08 | 0.75 ± 0.13 |
| XY | x = 75; y = 30; z = 0 | 2.60 ± 0.37 | 3.42 ± 0.42 | -0.82 ± 0.56 |
| YZ | x = 0; y = -30; z = 90 | 4.18 ± 0.09 | 2.62 ± 0.12 | 1.56 ± 0.15 |
| *XZ | x = 0; y = 55; z = 120 | 1.39 ± 0.04 | 0.95 ± 0.05 | 0.44 ± 0.06 |
| | | | | |
| *MRI-2 vs MRI-3* | | *MRI-2* | *MRI-3* | |
| XZ | x = -50; y = 0; z = -74 | 2.63 ± 0.08 | 2.16 ± 0.07 | 0.47 ± 0.11 |
| XY | x = -70; y = 45; z = 0 | 3.14 ± 0.50 | 3.38 ± 0.42 | -0.24 ± 0.65 |
| YZ | x = 0; y = 30; z = -95 | 2.72 ± 0.10 | 2.42 ± 0.13 | 0.30 ± 0.17 |
| *XZ | x = -45; y = 55; z = -70 | 1.59 ± 0.05 | 1.26 ± 0.04 | 0.33 ± 0.07 |

*relative to head/eye level

Tables 5 and 6 present the values for magnetic field and its spatial gradient, respectively, relative to three representative points (illustrated in Figure 15) at three distinct heights above the floor: 100 cm (y = - 5 cm), 130 cm (y = 25 cm), and 160 cm (y = 55 cm). For a person 170 cm tall, the chosen quotes represent approximately the quote of the waist, chest, and head/eyes [13].

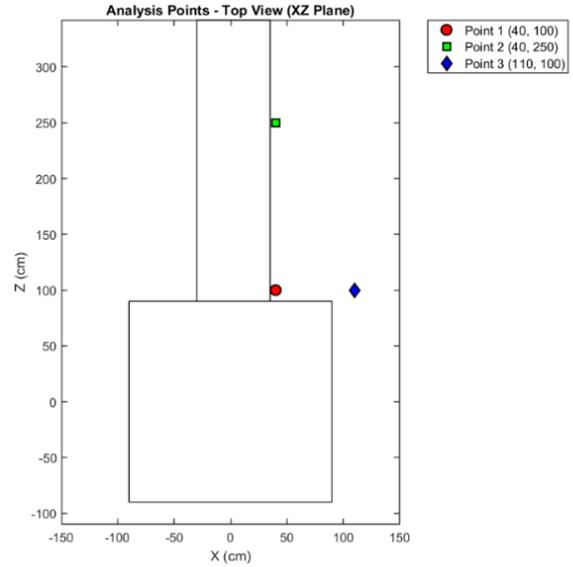

Fig. 15: Top view map of the room showing the three points taken as reference.

Table 5: Table showing the magnetic field differences for the three chosen points for the considered scanners (value ± uncertainty)

| Point | Height from the floor (cm) | Field Value (T) | Field Value (T) | Difference (T) |
|---|---|---|---|---|
| *MRI-1 vs MRI-2* | | *MRI-1* | *MRI-2* | |
| 1 (x = 40; z = 100; y = -5) | 100 | 1.01 ±0.025 | 0.93 ±0.023 | 0.08 ±0.034 |
| 1 (x = 40; z = 100; y = 25) | 130 | 0.88 ±0.022 | 0.78 ±0.019 | 0.10 ±0.029 |
| 1 (x = 40; z = 100; y = 55) | 160 | 0.48 ±0.012 | 0.45 ±0.011 | 0.03 ±0.016 |
| 2 (x = 40; z = 250; y = -5) | 100 | 0.01 ±0.00 | 0.00 ±0.00 | 0.01 ±0.00 |
| 2 (x = 40; z = 250; y = 25) | 130 | 0.01 ±0.00 | 0.00 ±0.00 | 0.01 ±0.00 |
| 2 (x = 40; z = 250; y = 55) | 160 | 0.00 ±0.00 | 0.00 ±0.00 | 0.00 ±0.00 |
| 3 (x = 110; z = 100; y = -5) | 100 | 0.17 ±0.007 | 0.16 ±0.004 | 0.01 ±0.008 |
| 3 (x = 110; z = 100; y = 25) | 130 | 0.14 ±0.006 | 0.13 ±0.003 | 0.01 ±0.007 |
| 3 (x = 110; z = 100; y = 55) | 160 | 0.07 ±0.003 | 0.08 ±0.002 | -0.01 ±0.004 |
| *MRI-1 vs MRI-3* | | *MRI-1* | *MRI-3* | |
| 1 (x = 40; z = 100; y = -5) | 100 | 1.01 ±0.025 | 0.99 ±0.027 | 0.02 ±0.037 |
| 1 (x = 40; z = 100; y = 25) | 130 | 0.88 ±0.022 | 0.84 ±0.022 | 0.04 ±0.031 |
| 1 (x = 40; z = 100; y = 55) | 160 | 0.48 ±0.012 | 0.49 ±0.013 | -0.01 ±0.017 |
| 2 (x = 40; z = 250; y = -5) | 100 | 0.01 ±0.00 | 0.00 ±0.00 | 0.01 ±0.00 |
| 2 (x = 40; z = 250; y = 25) | 130 | 0.01 ±0.00 | 0.00 ±0.00 | 0.01 ±0.00 |
| 2 (x = 40; z = 250; y = 55) | 160 | 0.00 ±0.00 | 0.00 ±0.00 | 0.00 ±0.00 |
| 3 (x = 110; z = 100; y = -5) | 100 | 0.17 ±0.007 | 0.22 ±0.006 | -0.05 ±0.009 |
| 3 (x = 110; z = 100; y = 25) | 130 | 0.14 ±0.006 | 0.18 ±0.004 | -0.04 ±0.007 |
| 3 (x = 110; z = 100; y = 55) | 160 | 0.07 ±0.003 | 0.11 ±0.003 | -0.04 ±0.004 |
| *MRI-2 vs MRI-3* | | *MRI-2* | *MRI-3* | |
| 1 (x = 40; z = 100; y = -5) | 100 | 0.93 ±0.023 | 0.99 ±0.027 | -0.06 ±0.035 |
| 1 (x = 40; z = 100; y = 25) | 130 | 0.78 ±0.019 | 0.84 ±0.022 | -0.06 ±0.029 |
| 1 (x = 40; z = 100; y = 55) | 160 | 0.45 ±0.011 | 0.49 ±0.014 | -0.04 ±0.017 |
| 2 (x = 40; z = 250; y = -5) | 100 | 0.00 ±0.00 | 0.00 ±0.00 | 0.00 ±0.00 |
| 2 (x = 40; z = 250; y = 25) | 130 | 0.00 ±0.00 | 0.00 ±0.00 | 0.00 ±0.00 |
| 2 (x = 40; z = 250; y = 55) | 160 | 0.00 ±0.00 | 0.00 ±0.00 | 0.00 ±0.00 |
| 3 (x = 110; z = 100; y = -5) | 100 | 0.16 ±0.004 | 0.22 ±0.006 | -0.06 ±0.007 |
| 3 (x = 110; z = 100; y = 25) | 130 | 0.13 ±0.003 | 0.18 ±0.004 | -0.05 ±0.005 |
| 3 (x = 110; z = 100; y = 55) | 160 | 0.08 ±0.002 | 0.11 ±0.003 | -0.03 ±0.003 |

Table 6: Table showing the magnetic gradient differences for the three chosen points for the considered scanners (value ± uncertainty)

| Point | Height from the floor (cm) | Gradient Magnitude (T/m) | Gradient Magnitude (T/m) | Difference (T/m) |
|---|---|---|---|---|
| *MRI-1 vs MRI-2* | | *MRI-1* | *MRI-2* | |
| 1 (x = 40; z = 100; y = -5) | 100 | 3.30 ±0.08 | 2.72 ±0.07 | 0.58 ±0.10 |
| 1 (x = 40; z = 100; y = 25) | 130 | 3.50 ±0.08 | 2.79 ±0.06 | 0.71 ±0.10 |
| 1 (x = 40; z = 100; y = 55) | 160 | 2.18 ±0.04 | 1.98 ±0.04 | 0.20 ±0.06 |
| 2 (x = 40; z = 250; y = -5) | 100 | 0.03 ±0.00 | 0.01 ±0.00 | 0.02 ±0.00 |
| 2 (x = 40; z = 250; y = 25) | 130 | 0.02 ±0.00 | 0.00 ±0.00 | 0.02 ±0.00 |
| 2 (x = 40; z = 250; y = 55) | 160 | 0.02 ±0.00 | 0.00 ±0.00 | 0.02 ±0.00 |
| 3 (x = 110; z = 100; y = -5) | 100 | 1.11 ±0.05 | 0.74 ±0.02 | 0.37 ±0.05 |
| 3 (x = 110; z = 100; y = 25) | 130 | 0.80 ±0.04 | 0.67 ±0.02 | 0.13 ±0.04 |
| 3 (x = 110; z = 100; y = 55) | 160 | 0.57 ±0.02 | 0.36 ±0.01 | 0.21 ±0.02 |
| *MRI-1 vs MRI-3* | | *MRI-1* | *MRI-3* | |
| 1 (x = 40; z = 100; y = -5) | 100 | 3.30 ±0.08 | 2.51 ±0.07 | 0.79 ±0.10 |
| 1 (x = 40; z = 100; y = 25) | 130 | 3.50 ±0.08 | 2.61 ±0.07 | 0.89 ±0.11 |
| 1 (x = 40; z = 100; y = 55) | 160 | 2.18 ±0.04 | 2.13 ±0.05 | 0.05 ±0.06 |
| 2 (x = 40; z = 250; y = -5) | 100 | 0.03 ±0.00 | 0.01 ±0.00 | 0.02 ±0.00 |
| 2 (x = 40; z = 250; y = 25) | 130 | 0.02 ±0.00 | 0.00 ±0.00 | 0.02 ±0.00 |
| 2 (x = 40; z = 250; y = 55) | 160 | 0.02 ±0.00 | 0.00 ±0.00 | 0.02 ±0.00 |
| 3 (x = 110; z = 100; y = -5) | 100 | 1.11 ±0.05 | 1.25 ±0.03 | -0.13 ±0.05 |
| 3 (x = 110; z = 100; y = 25) | 130 | 0.80 ±0.04 | 0.99 ±0.03 | -0.19 ±0.05 |
| 3 (x = 110; z = 100; y = 55) | 160 | 0.57 ±0.02 | 0.58 ±0.01 | -0.01 ±0.02 |
| *MRI-2 vs MRI-3* | | *MRI-2* | *MRI-3* | |
| 1 (x = 40; z = 100; y = -5) | 100 | 2.72 ±0.07 | 2.51 ±0.07 | 0.21 ±0.09 |
| 1 (x = 40; z = 100; y = 25) | 130 | 2.79 ±0.06 | 2.61 ±0.07 | 0.18 ±0.09 |
| 1 (x = 40; z = 100; y = 55) | 160 | 1.98 ±0.04 | 2.13 ±0.05 | -0.15 ±0.06 |
| 2 (x = 40; z = 250; y = -5) | 100 | 0.01 ±0.00 | 0.01 ±0.00 | -0.00 ±0.00 |
| 2 (x = 40; z = 250; y = 25) | 130 | 0.00 ±0.00 | 0.00 ±0.00 | -0.00 ±0.00 |
| 2 (x = 40; z = 250; y = 55) | 160 | 0.00 ±0.00 | 0.00 ±0.00 | -0.00 ±0.00 |
| 3 (x = 110; z = 100; y = -5) | 100 | 0.74 ±0.02 | 1.25 ±0.03 | -0.51 ±0.04 |
| 3 (x = 110; z = 100; y = 25) | 130 | 0.67 ±0.01 | 0.99 ±0.03 | -0.32 ±0.03 |
| 3 (x = 110; z = 100; y = 55) | 160 | 0.36 ±0.01 | 0.58 ±0.01 | -0.22 ±0.01 |

## 5 Discussion

This study is principally dedicated to analysing and measuring the differences in magnetic field distributions produced by 3 Tesla MRI systems across various clinical settings. By emphasizing magnetic field leakage, the research aims to contribute meaningful data to the domain of MRI safety, enhancing the understanding of environmental magnetic exposure. The findings will serve as a critical resource for healthcare facilities looking to improve room configuration and implement effective safety strategies, especially in response to the rising adoption of high-field MRI technologies.

The present study utilized experimental measurements of the fringe field around three MRI scanners situated in three distinct clinical facilities, employing a home-made script to obtain two-dimensional maps of the fringe field and its gradient on the axial, sagittal, and coronal planes through the isocentre.

The implementation of the software is based on a modular architecture, which is comprised of two distinct phases. The initial phase is responsible for the management of dimensional data from the MRI suite, incorporating both the characteristic parameters of the room and the interpolated planes. The subsequent phase employs a three-dimensional interpolation function that is generalised in order to generate the complete map, based on the preprocessed data.

This computational framework facilitates flexible and scalable management of different MRI installations. The specification of the reference room is all that is required to generate the corresponding complete field mapping. The interpolation methodology described herein is generalisable to other MRI suites, provided that experimental measurements encompass a minimum surface area of

70 × 70 cm along the x and z axes. This methodology affords high-resolution representations of magnetic field distributions, thereby ensuring precise depiction of the spatial characteristics of fringe fields within the context of clinical MRI environments.

A close examination of the results obtained reveals the following observations:

1. With regard to the magnetic field, it can be observed that there is a maximum difference of 0.5 T between scanners from different manufacturers, while the maximum difference is reduced to 0.17 T when comparing the two scanners from the same manufacturer. In both cases, the maximum observed difference was located at the edge of the gantry: in the first case, in the central area with respect to the patient bed, in the second case, in a more lateral area. These zones correspond to the primary working positions of technicians during patient setup, thereby subjecting them to full-body exposure to the magnetic field. In accordance with the initial assumptions, the results obtained are deemed equally valid for areas that are symmetrical with respect to the isocentre.
2. With regard to the spatial gradient, it can be observed that there is a maximum difference of 1.64 T/m between scanners from different manufacturers, while the maximum difference is reduced to 0.5 T/m when comparing the two scanners from the same manufacturer. As demonstrated in Figures 11 and 12, there is a significant disparity in the spatial gradient values of two scanners from different manufacturers in the anterior/posterior (both XZ and YZ planes) area in proximity to the gantry.
3. As shown by Tables 4 and 5, the differences for the magnetic field and its spatial gradient are highly dependent on the location in the space around the MRI scanner. In an effort to generalise, it can be stated that the most significant discrepancies in absolute terms are located in close proximity to the gantry. By fixing the position of a specific point with respect to the x and z axes, it is possible to observe a decrease in the differences in magnetic field and spatial gradient as the distance from the floor increases.

The final outcomes suggest that the fringe field and spatial gradient difference between scanners of different manufacturers may be attributable to a different type of shielding [14], [15]. Additionally, although less pronounced, discrepancies in magnetic field values and its spatial gradient were also identified between scanners of the same manufacturer.

In summary, as revealed by our findings, the fringe field and spatial gradient differed across all sites, despite the scanners having the same nominal magnetic field strength ($B_0$ = 3 T).

It is widely acknowledged that MRI operators undergo training to familiarise them with the potential biological effects that may arise from their work at MRI sites. However, they lack a subjective perception of the actual distribution of the magnetic field and its spatial gradient in their specific work environment. The findings of this study demonstrate the necessity of acquiring precise information regarding the fringe field and spatial gradient in the context of specific MRI environments. The generic information provided by manufacturers is insufficient in addressing the potential for substantial variations in exposure, which can occur even between scanners of the same brand. A specific analysis permits the acquisition of information pertaining to the areas within the MRI facility where personnel are susceptible to elevated exposure levels. The comprehension of magnetic field and its spatial gradient measurement provides indications regarding high-risk areas, in which the operator should exercise caution and proceed with extreme caution. These recommendations are of particular significance for those responsible for the organisation of the workplace and work practice.

The present study acknowledges certain limitations, primarily pertaining to the uncertainty surrounding the employed measurement method and analytical model. In order to enhance the robustness of the findings, it is advisable that further extensive validation is carried out through simulations or repeated independent measurements. However, conducting simulations of different MRI scanner configurations – such as altered shielding, room geometries, or scanner orientations – would require detailed technical specifications and proprietary information from manufacturers, which are typically not publicly available. A further limitation pertains to the model utilised in generating the fringe field map encompassing the scanner. This model operates under the assumption of field symmetry with respect to the isocentre. Nevertheless, this assumption may not hold true due to the distinct configuration of the room and its structural characteristics. In conclusion, the study is based on a limited number of MRI scanners and sites, which restricts its ability to draw broader conclusions about scanner variability or magnetic shielding effects across different manufacturers or environments.

In conclusion, this study may be regarded as a preliminary step towards more comprehensive evaluations that may include operator movement and time-dependent field variation.

**Contribution of Individual Authors to the Creation of a Scientific Article (Ghostwriting Policy)**
FG, VH: Conceptualization, Data curation, Investigation, Methodology, Software, Visualization, Writing – original draft, Writing – review and editing. VMF: Data curation, Writing – original draft. MM, MDA: Project administration, Resources, Writing – review and editing VH, GA: Conceptualization, Formal Analysis, Funding acquisition, Project administration, Supervision, Writing – review and editing.

**Sources of Funding for Research Presented in a Scientific Article or Scientific Article Itself**
This research was funded by INAIL (National Institute for Insurance against Accidents at Work), grant number Bric 2022 CUP: J43C22001390005.

**Conflict of Interest**
The authors have no conflicts of interest to declare that are relevant to the content of this article.